# Pressure-induced phase transitions of halogen-bridged binuclear metal complexes $R_4[Pt_2(P_2O_5H_2)_4X]\cdot nH_2O$

Shoji Yamamoto

*Department of Physics, Okayama University, Tsushima, Okayama 700-8530, Japan*

Recent contrasting observations for halogen ($X$)-bridged binuclear platinum complexes $R_4[Pt_2(P_2O_5H_2)_4X]\cdot nH_2O$, that is, pressure-induced Peierls and reverse Peierls instabilities, are explained by finite-temperature Hartree-Fock calculations. It is demonstrated that increasing pressure transforms the initial charge-polarization state into a charge-density-wave state at high temperatures, whereas the charge-density-wave state oppositely declines with increasing pressure at low temperatures. We further predict that higher-pressure experiments should reveal successive phase transitions around room temperature.

Quasi-one-dimensional halogen ($X$)-bridged metal ($M$) complexes [1], which are referred to as $MX$ chains, have been attracting much interest for several decades. They present an interesting stage performed by electron-electron correlation, electron-lattice interaction, low dimensionality, and $d$-$p$ orbital hybridization [2]. The Mott and Peierls insulators compete with each other in their ground states, while topological defects such as solitons and polarons appear in their excited states exhibiting unique transport properties [3]. In recent years, a new class of these materials [4–7], which is characterized by binuclear metal complexes bridged by halogens and is thus referred to as $MMX$ chains, has stimulating further interest in one-dimensional unit-assembled spin-charge-lattice coupling systems. In comparison with $MX$ chains, $MMX$ chains indeed possess fascinating features. The formal oxidation state of the metal ions is $3+$ in $MX$ chains, whereas it is $2.5+$ in $MMX$ chains. Therefore, $MMX$ chains have an unpaired electron per metal dimer even in their trapped-valence states, contrasting with the valence-trapped state consisting of $M^{2+}$ and $M^{4+}$ in $MX$ chains. The $M(d_{z^2})$-$M(d_{z^2})$ direct overlap in $MMX$ chains effectively reduces the on-site Coulomb repulsion due to its $d_{\sigma^*}$ character and therefore enhances the electrical conductivity. Some of $MMX$ compounds [4,5] have a neutral chain structure, where the metal sublattice gets rid of the hydrogen-bond network, and thus exhibit more pronounced one dimensionality.

Four possible one-dimensional charge-ordering modes of $MMX$ chains were empirically pointed out [8–10] and they were indeed verified theoretically [11,12]. Potential magnetic phases have also been proposed [13] and a systematic study of broken-symmetry solutions in the two-band scheme has been presented [13,14]. Let us introduce the $\frac{5}{6}$-filled one-dimensional two-band three-orbital extended Peierls-Hubbard Hamiltonian:

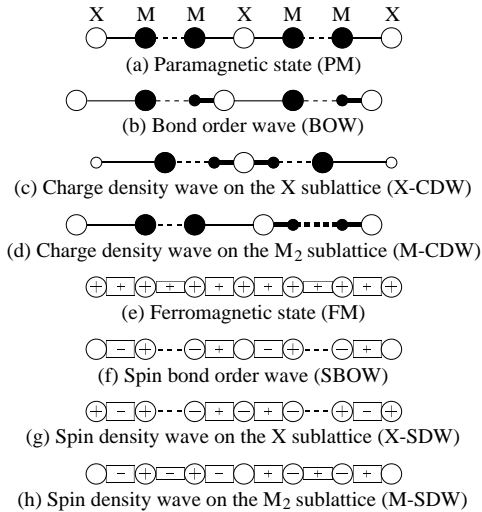

FIG. 1. Schematic representation of possible density-wave states. The various circles and segments qualitatively denote the variation of local electron densities and bond orders, respectively, whereas the signs ± in circles and strips denote the alternation of local spin densities and spin bond orders, respectively.

$$\mathcal{H} = \sum_{m,s}\sum_{j=1,2} (\varepsilon_M - \beta l_{j:n}) n_{j:m,s} + \sum_{m,s}\varepsilon_X n_{3:m,s}$$
$$- \sum_{m,s}\sum_{j=1,2} (t_{MX} - \alpha l_{j:n})\bigl(a^\dagger_{j:n,s}a_{3:n,s} + a^\dagger_{3:n,s}a_{j:n,s}\bigr)$$
$$- \sum_{m,s} t_{MM}\bigl(a^\dagger_{1:n,s}a_{2:n-1,s} + a^\dagger_{2:n-1,s}a_{1:n,s}\bigr)$$
$$+ \sum_{m}\sum_{j=1,2} U_M\, n_{j:m,+}n_{1:m,-} + \sum_m U_X\, n_{3:n,+}n_{3:n,-}$$
$$+ \sum_{m,s,s'}\sum_{j=1,2} V_{MX}\, n_{j:m,s}n_{3:m,s'}$$
$$+ \sum_{m,s,s'} V_{MM}\, n_{1:m,s}n_{2:m-1,s'} + \sum_m\sum_{j=1,2}\frac{K}{2}l_{j:m}^2, \quad (1)$$

where $n_{j:m,s} = a^\dagger_{j:m,s}a_{j:m,s}$ with $a^\dagger_{j:m,s}$ being the creation operator of an electron with spin $s = \pm$ (up and down) for the $M$ $d_{z^2}$ ($j=1,2$) or $X$ $p_z$ ($j=3$) orbital in the $m$th $MXM$ unit, and $l_{j:m} = (-1)^j(u_{j:m} - u_{3:m})$ with $u_{j:m}$ being the chain-direction displacement of the metal ($j = 1, 2$) or halogen ($j = 3$) in the $m$th $MXM$ unit from its equilibrium position. $\alpha$ and $\beta$ are, respectively, the intersite and intrasite electron-lattice coupling constants, while $K$ is the metal-halogen spring constant. We assume, based on the thus-far reported experimental observations, that every $M_2$ moiety is not deformed, namely, $u_{1:n} = u_{2:n-1}$. $\varepsilon_M$ and $\varepsilon_X$ are the on-site energies of isolated metal and halogen atoms, respectively. The electron hoppings between these levels are modeled by $t_{MM}$ and $t_{MX}$, whereas the electron-electron Coulomb interactions by $U_M$, $U_X$, $V_{MM}$, and $V_{MX}$. We always set $t_{MM}$ and $K$ both equal to unity. This Hamiltonian possesses various density-wave solutions [13], which are schematically shown in Fig. 1. (b) to (d) are accompanied by lattice distortion, where the latter two exhibit cell doubling. (e) to (h) are possible spin alignments and their nonlocal stabilization assumes weak interchain interaction. (c) and (g) are more characterized by charge and spin modulation on the $X$ sublattice, respectively, than by any density wave on the $M_2$ sublattice. The relevance of the $X$ $p_z$ orbitals to these states has fully been demonstrated [13].

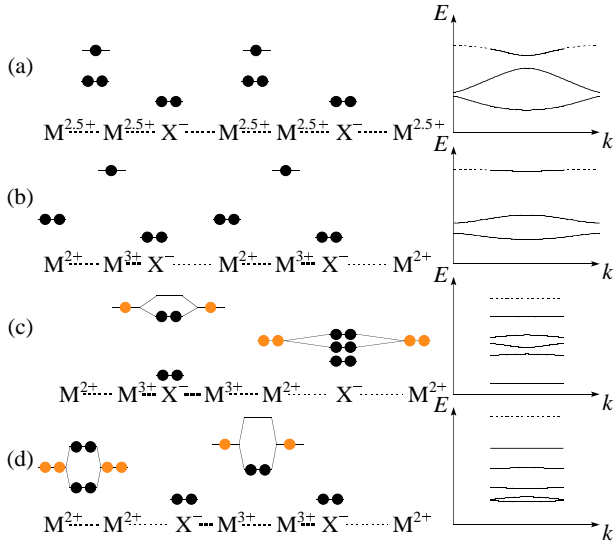

FIG. 2. Local electronic structures and the resultant band structures of the four charge-ordering modes provided the Coulomb interaction is not so strong, where the valence numbers denote formal oxidation states, namely, 2+ and 3+ should generally be regarded as $(2 + \delta)+$ and $(3 - \delta)+$ ($0 \leq \delta \leq 0.5$), respectively: (a) PM, (b) BOW, (c) X-CDW, and (d) M-CDW.

The existent $MMX$ complexes consist of two families: $M_2(dta)_4I$ ($M = Pt, Ni$; dta = dithioacetate = $CH_3CS_2^-$) and $R_4[Pt_2(pop)_4X] \cdot nH_2O$ ($X = Cl, Br, I$; $R = Li, K, Cs, \cdots$; pop = diphosphonate = $P_2O_5H_2^{2-}$). The former, dta complexes, are not yet well investigated and one of them, $Pt_2(dta)_4I$, has just come into hot argument due to recent stimulative observations [15]: With decreasing temperature, there occurs a metal-semiconductor transition, that is, a transition from the averaged-valence state (a) to the trapped-valence state (b), at 300 K, and further transition to the charge-ordering mode (c) follows around 80 K. The room-temperature conductivity is larger by nine digits than those of typical $MX$ chains, while the metal-sublattice dimerization has never been observed in $MX$ chains. On the other hand, the latter, pop complexes, have extensively been measured [8,9] and their ground states have generally been assigned to (d). However, due to the small Peierls gap, the ground states of these materials can be tuned by replacing the halogens and/or counter ions [16]. Such a tuning of the electronic state can be realized also by pressure [17] and the pressure-induced phase transitions of pop complexes are the central issue in this article.

Figure 2 shows that the orbital hybridization within every $M_2$ moiety, which depends on the electron transfer $t_{MM}$, is essential to the stabilization of M-CDW, whereas X-CDW owes its stabilization to the interdimer electron transfer and thus to $t_{MX}$. As the adjacent metals in the dimer are tightly locked to each other by the surrounding ligands, an applied pressure mainly reduces the $M$-$X$ distance. Therefore, from the viewpoint of the electron transfer, increasing pressure should stabilize X-CDW instead of M-CDW. Swanson et al. [18] indeed observed a *pressure-induced reverse Peierls instability*, namely, the disappearance of the two sharp bands originating from the $Pt^{2+}$-$Pt^{2+}$ and $Pt^{3+}$-$Pt^{3+}$ stretching modes in the Raman spectra and the red shift of the intervalence charge-transfer band due to the $d_{\sigma^*}(Pt^{2+}) \to d_{\sigma^*}(Pt^{3+})$ charge-transfer gap in the absorption spectra with increasing pressure, for a pop-family $MMX$ compound $K_4[Pt_2(pop)_4Br] \cdot 3H_2O$ at 20 K. However, there has quite recently appeared another report [19] that an applied pressure stabilizes M-CDW. Matsuzaki et al. have systematically synthesized numerous pop-family iodo complexes changing the counter ions and have found that as the $M$-$X$-$M$ distance increases, their valence structure is generally tuned from (d) to (b). They applied pressure to one of these samples, $[(C_2H_5)_2NH_2]_4[Pt_2(pop)_4I]$, at room temperature and observed a phase transition from BOW to M-CDW, that is, a *pressure-induced Peierls instability*. Increasing interdimer charge transfer and resultant charge proportionation look disadvantageous to M-CDW as well as to BOW. Neither the $M$-$X$ Coulomb repulsion $V_{MX}$ nor the interdimer direct one $V_{MXM}$ distinguishes between these states within the single-band scheme of the electronic configuration (Fig. 2). There must lie intertwining driving forces in the BOW-to-M-CDW transition. It is our aim to explain the two contrasting pressure-induced phenomena consistently.

We calculate the free energy for the broken-symmetry solutions (a) to (h) of the Hamiltonian (1) within the Hartree-Fock approximation. The lattice distortion is adiabatically determined at each temperature so as to minimize the free energy. Although the Hamiltonian (1) have numerous parameters in itself, there is still plenty which may be taken into the model. For example, the halogen on-site energy may also be coupled to lattice, while there may be an interdimer direct spring constant due to the hydrogen bonds between the ligands and counter ions. However, a weak alternation of the lower orbital energy is much less important and the coexistent elastic forces can be renormalized into $\alpha$ and $\beta$. Indeed, the present model can successfully reproduce the observed optical conductivity [20], suggesting that all the essential interactions can well be described by the parameters on hand. In such circumstances that even the hopping amplitudes are not yet determined, we should minimize the number of parameters.

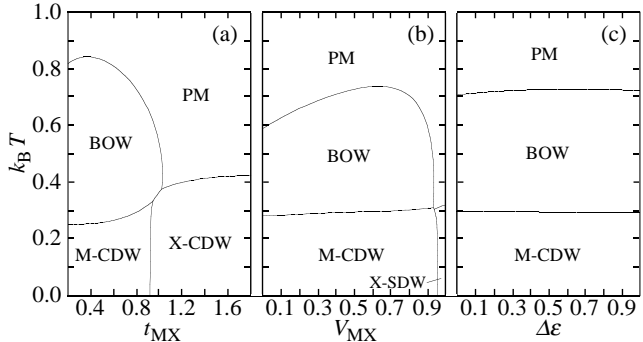

FIG. 3. Thermal phase boundaries as functions of $t_{MX}$ (a), $V_{MX}$ (b), and $\Delta\varepsilon$ (c), where the parametrization, $\alpha = 0.7$, $\beta = 1.4$, $\Delta\varepsilon = 0.5$, $t_{MX} = 0.8$, $U_M = 1.5$, $U_X = 1.0$, $V_{MM} = 0.5$, and $V_{MX} = 0.5$, is common to all except for the variable.

We show in Fig. 3 typical thermal behaviors as functions of $t_{MX}$ and $V_{MX}$, both of which are expected to increase under an applied pressure. According to expectation, Fig. 3(a) shows that M-CDW is transformed into X-CDW as $t_{MX}$ increases. The idea of interdimer charge proportionation with increasing $t_{MX}$ is still applicable to the high-temperature behavior. At low temperatures, the electrons lower their energy with the Peierls gap open, but with increasing temperature, the entropy becomes relevant to the free energy. Since its contribution is expressed as $TS = \sum_k \{(E_k - \mu)f_k + k_B T \ln[1 + e^{-(E_k-\mu)/k_B T}]\}$, where $f_k = [1 + e^{(E_k-\mu)/k_B T}]^{-1}$ with the energy dispersion $E_k$ and the chemical potential $\mu$, the electrons come to prefer the metallic states BOW and PM to the Peierls-gap states M-CDW and X-CDW with increasing temperature. BOW exhibits charge disproportionation within each $M$-$X$-$M$ moiety and it is thus replaced by PM with increasing $t_{MX}$. In a thermal stabilization, BOW precedes PM due to its flatter conduction band (Fig. 2), which is more advantageous for gaining the entropy. As the thermal phase boundary between BOW and M-CDW weakly depends on $t_{MX}$, BOW may in principle be transformed into M-CDW with increasing $t_{MX}$. However, such a transition is possible only in a narrow temperature range. Then we inquire into the effect of the electronic correlation. We focus on the thermal behavior at $t_{MX} = 0.8$ in Fig. 3(a) and visualize its dependence on $V_{MX}$ in Fig. 3(b). Though a significant increase of $V_{MX}$ destabilizes M-CDW and leads to the oxidation of the halogen ions inducing spin moments on the halogen sites, the competition between BOW and M-CDW looks much less sensitive to $V_{MX}$ varying independently of $t_{MX}$.

Besides the electron transfer and correlation, the onsite electron affinity may be influenced by an applied pressure. On the one hand, increasing pressure enhances the overlap between the adjacent $M$ $d_{z^2}$ and $X$ $p_z$ orbitals, effectively leading to a reduction of the difference between the two orbital energies, $\Delta\varepsilon = \varepsilon_M - \varepsilon_X$. On the other hand, from the viewpoint of the Madelung energy, a reduction of the $M$-$X$ distance causes the Coulomb potential on the metal sites to rise and therefore enhances $\Delta\varepsilon$. Hence it is likely that the resultant net change of $\Delta\varepsilon$ is not so significant as those of $t_{MX}$ and $V_{MX}$. Furthermore, the electronic state turns out insensitive to $\Delta\varepsilon$ under a reasonable condition $t_{MX} \lesssim t_{MM}$ [6,7]. The thermal behavior at $t_{MX} = 0.8$ in Fig. 3(a) is again taken up and its dependence on $\Delta\varepsilon$ is shown in Fig. 3(c). The electronic state shows little dependence on $\Delta\varepsilon$ all over the temperature range. Thus, we reasonably assume that an applied pressure only influences the $M$-$X$ electron transfer and Coulomb repulsion.

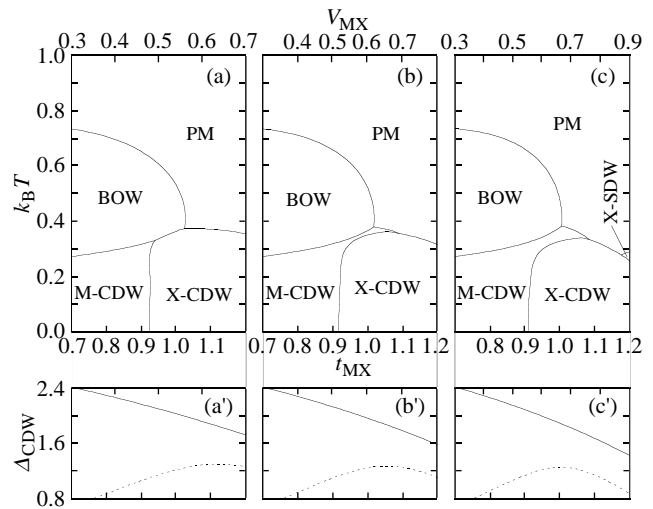

FIG. 4. Thermal phase boundaries and the Peierls gaps $\Delta_{CDW}$ as functions of $t_{MX}$ and $V_{MX}$ varying under connections $\Delta V_{MX} = 0.8 \Delta t_{MX}$ (a, a'), $\Delta V_{MX} = 1.0 \Delta t_{MX}$ (b, b'), and $\Delta V_{MX} = 1.2 \Delta t_{MX}$ (c, c'), where the rest of the parameters are the same as those in Fig. 3. In (a') to (c'), the solid and dotted lines represent the gaps for M-CDW and X-CDW, respectively.

Now we are eager to look into the combined pressure effects. In Fig. 4 we observe the electronic state connecting the increase of the Coulomb repulsion, $\Delta V_{MX}$, with the enhancement of the electron transfer, $\Delta t_{MX}$.

The low-temperature reverse Peierls instability with increasing pressure is steadfast under any parametrization. With increasing pressure, the energy gap decreases and then increases via the first-order transition from M-CDW to X-CDW, though the latter process has not yet been observed explicitly [21]. On the other hand, the high-temperature behavior is much more sensitive to the parametrization. Assuming a predominant pressure effect on $t_{MX}$, there exists a little possibility of BOW being transformed into M-CDW. BOW is more likely to change into X-CDW at intermediate temperatures and into PM at sufficiently high temperatures with increasing pressure. As we switch on the other pressure effect on $V_{MX}$, M-CDW begins to grow between BOW and X-CDW. Although this is no longer a simple quantum competition, it may still be useful to observe the resultant changes of the Peierls gaps. We ascribe the pressure-induced BOW-to-M-CDW transition to a thermal competition between M-CDW and X-CDW rather than to any competition between BOW and M-CDW themselves, because growing M-CDW appears to go with diminishing X-CDW in Fig. 4. Increasing $t_{MX}$ destabilizes M-CDW and stabilizes X-CDW, reducing and enhancing their quantum energy gaps. Within a single-band description assuming the $X$ $p_z$ orbitals to be stably filled, there occurs no quantum competition due to $V_{MX}$ between M-CDW and X-CDW. Once we consider the oxidation of the halogen ions and its contribution to the stabilization of X-CDW [13], increasing $V_{MX}$ more advantageously acts on X-CDW than on M-CDW and indeed accelerates the increase and decrease of their energy gaps (Figs. 4(a′) to 4(c′)). At finite temperatures, decreasing gap induces thermal fluctuations and contributes to gaining the entropy. We are thus convinced that simultaneously increasing $t_{MX}$ and $V_{MX}$ more stabilize X-CDW than M-CDW in the quantum competition, while vice versa in the thermal competition.

Thus, we are led to the conclusion that the low-temperature pressure-induced reverse Peierls instability [18] is essentially ascribed to the resultant enhancement of the electron transfer, whereas the combined increases of the electron transfer and the Coulomb repulsion are really relevant to the room-temperature pressure-induced Peierls instability [19], where the latter effect predominates over the former one in magnitude. If the present scenario is true, successive phase transitions are expected with increasing pressure in a certain temperature range. The low-temperature pressure effects on $K_4[Pt_2(pop)_4X] \cdot 3H_2O$ ($X = $ Cl, Br) [18,21] were measured up to 10 GPa, while the room-temperature ones on $[(C_2H_5)_2NH_2]_4[Pt_2(pop)_4I]$ [19] within 2 GPa. Therefore, the latter experiments can further be developed. Higher-pressure experiments around room temperature may reveal successive transitions such as BOW → M-CDW → X-CDW and BOW → M-CDW → PM. Such observations may become more feasible by tuning the initial electronic state toward the BOW-M-CDW phase boundary, for example, with the replacement of the counter ions by $NH_3(C_3H_6)NH_3$ or $(C_5H_{11})_2NH_2$ [19]. We stress the direct transition from M-CDW to PM as an evidence of the predominant pressure effect on the Coulomb interaction. No pop-family *MMX* compound whose ground state is X-CDW has been found so far, but it may exist under pressure. On the other hand, the dta-family platinum complex $Pt_2(dta)_4I$ exhibits a ground state of the X-CDW type and must display different pressure-induced phenomena. Further extensive measurements on both pop- and dta-family *MMX* compounds under high static pressure are encouraged.

The author is grateful to Prof. K. Yonemitsu for fruitful discussion. He further thanks Prof. H. Okamoto for a communication on his experimental findings prior to publication. This work was supported by the Japanese Ministry of Education, Science, Sports, and Culture. The numerical calculation was done using the facility of the Supercomputer Center, Institute for Solid State Physics, University of Tokyo.